\documentclass{article}
\usepackage{blindtext}
\usepackage[a4paper, total={7in, 8in}]{geometry}
\usepackage[english]{babel}
\usepackage{amsmath}
\usepackage{makecell}
\usepackage{booktabs}%\toprule、\midrule、\bottomrule
\usepackage{multirow}
\usepackage{float}
\usepackage{graphicx}
 \usepackage{cite}
\usepackage[colorlinks=true, allcolors=blue]{hyperref}
\usepackage{algorithm}
\usepackage{color}

\providecommand{\keywords}[1]
{
  \small	
  \textbf{\textit{Keywords---}} #1
}
\begin{document}
%\title{3D Ultrasound Localisation Microscopy Using Online 4D Ultrasound-Based Robotic Tracking for Large Tissue Displacements}
\title{Online 4D Ultrasound-Guided Robotic Tracking Enables 3D Ultrasound Localisation Microscopy with Large Tissue Displacements}
\author{Jipeng Yan, Qingyuan Tan, Shusei Kawara,  Jingwen Zhu, Bingxue Wang,  \\
Matthieu Toulemonde, 
Honghai Liu, Ying Tan, Meng-Xing Tang
\thanks{This work was supported by the Engineering and Physical Sciences Research Council under Grant No. EP/X033651/1 and EP/V04799X/1, and the National Institute for Health Research i4i under Grant No. NIHR200972. (\emph{Corresponding authors: Meng-Xing Tang, email: mengxing.tang @imperial.ac.uk})
\\
Jipeng was with Ultrasound Lab for Imaging and Sensing, Department of Bioengineering, Imperial College London, London SW7 2AZ, UK, and is with the State Key Laboratory of Robotics and Systems, Harbin Institute of Technology, Harbin 150001, China. (e-mail: jipengyan@hit.edu.cn)
\\ Qingyuan, Jingwen, Bingxue, Matthieu and Meng-Xing are with Ultrasound Lab for Imaging and Sensing, Department of Bioengineering, Imperial College London, London SW7 2AZ, UK. (e-mail: l.tan21, jingwen.zhu22, bingxue.wang18,  m.toulemonde, mengxing.tang, @imperial.ac.uk
\\
Shusei is with the Department of Bioengineering, Imperial College London, London, SW7 2AZ, UK. (e-mail: s.kawara19@imperial.ac.uk).
\\
Honghai is with the School of Computing, University of Portsmouth, Portsmouth, PO1 2UP
UK. (e-mail: honghai.liu@port.ac.uk). 
\\
Ying is with the Department of Mechanical Engineering,
University of Melbourne, Parkville, VIC 3010, Australia (e-mail:
yingt@unimelb.edu.au
}
}
\date{}
\maketitle
\begin{abstract}
Super-Resolution Ultrasound (SRUS) imaging through localising and tracking microbubbles, also known as Ultrasound Localisation Microscopy (ULM), has demonstrated significant potential for reconstructing microvasculature and flows with sub-diffraction resolution in clinical diagnostics. However, imaging organs with large tissue movements, such as those caused by respiration, presents substantial challenges. Existing methods often require breath holding to maintain accumulation accuracy, which limits data acquisition time and ULM image saturation. To improve image quality in the presence of large tissue movements, this study introduces an approach integrating high-frame-rate ultrasound with online precise robotic probe control. Tested on a microvasculature phantom with translation motions up to 20 mm, twice the aperture size of the matrix array used,  our method achieved real-time tracking of the moving phantom and imaging volume rate at 85 Hz, keeping majority of the target volume in the imaging field of view. ULM images of the moving cross channels in the phantom were successfully reconstructed in post-processing, demonstrating the feasibility of super-resolution imaging under large tissue motions. This represents a significant step towards ULM imaging of organs with large motion.
\end{abstract}


\section{Introduction}

Super-Resolution Ultrasound (SRUS) imaging, based on the localisation \cite{Couture2011MUSLI} and tracking \cite{Siepmann2011ImagingTumor} of sparse microbubbles (MBs) \cite{christensen2020super,song2023super}, also known as Ultrasound Localisation Microscopy (ULM) \cite{couture2018ultrasound,dencks2023ultrasound}, has been validated for reconstructing microvasculature and flows both \textit{in vitro} \cite{viessmann2013acoustic,desailly2013sono} and in animal models \cite{christensen2014vivo,errico2015ultrafast,song2017improved,lin20173,zhu20193d,qian2020vivo,qiu2022vivo,taghavi2022ultrasound,demeulenaere2022coronary}, showing promise for clinical applications in organs such as breast \cite{opacic2018motion}, liver \cite{huang2021super},  brain \cite{demene2021transcranial}, lymph node \cite{zhu2022super}, kidney \cite{huang2021super,denis2023sensing,bodard2024visualization}, testis \cite{li2024super}, and heart \cite{yan2024transthoracic}.

%(breath holding limited image quality) 

Although ULM achieves unprecedented high spatial resolution by accumulating localised MBs, it faces some significant technical challenges, particularly tissue motions. These tissue movements can blur the reconstructed vasculature and compromise the accuracy of hemodynamic measurements, or even worse, generate ghost vessels\cite{harput2018two}. Tissues may be displaced by large vessels with pulsatility or by the lungs during respiration. Compared to the pulsatility of large vessels, respiratory-induced tissue motions are considerably larger. For example, human livers have been reported to shift up to 20 mm due to respiration \cite{fahmi2018respiratory}. Consequently, while ULM performs well for organs less affected by the lungs—such as the human brain \cite{demene2021transcranial} or transplanted kidneys \cite{denis2023sensing}—its effectiveness decreases for organs moving significantly during breathing, such as native kidneys \cite{bodard2024visualization}, liver \cite{huang2021super}, and heart \cite{yan2024transthoracic}. In these cases, ULM is generally limited to the duration of a breath hold. As a result, ULM images of such organs often exhibit reduced saturation compared to those of brain images, which limits resolution and affects the reliability of quantification \cite{christensen2019poisson, hingot2019microvascular, dencks2020assessing}.

%(existing post-processing techniques make breath holding necessary for acquisition) 

There are multiple reasons necessitating breath holding during ULM imaging. MBs can move out of the imaging field of view (FoV) when breathing. Existing methods cannot recover contrast signals under such situations. Additionally, motion correction techniques do not always perform satisfactorily, especially in the presence of significant tissue motion when speckles significantly decorrelate. For instance, Doppler-based motion detection methods are not effective enough for detecting motions along the lateral direction, particularly for deep tissues \cite{cormier2021dynamic, poree2016high}. Image registration methods often face convergence issues, frequently getting trapped in local minima, particularly when initialisations for optimisation are far from the global minimum. This issue is exacerbated by the variability in tissue motion across spatial, temporal, and subject conditions \cite{sotiras2013deformable, harput2018two}. Moreover, the point spread function (PSF) varies across the ultrasound view, causing changes in the image pattern of fixed structures based on their position, which further impacts the accuracy of image registration and speckle tracking \cite{taghavi2021vivo}. Consequently, to ensure that ULM images are reconstructed with high resolution using existing motion correction methods, patients are typically required to hold their breath to prevent the target vasculature from moving out of the imaging view \cite{song2023super}.

%(why using robot and related background) 

Eliminating the need for breath holding during ULM data acquisition requires substantial advancements in post-processing techniques, such as learning-based methods \cite{jiang2021review}, to handle information loss and compensate for large motions in the acquired data. Triggering breathing phase with data acquisition might help reduce the challenge in motion correction, but only part of data is useful for the accumulation in a region of interest. An alternative approach involves moving ultrasound probes in tandem with the target during respiration to keep it within the imaging FoV throughout data acquisition. This can be formulated as a tracking problem. While human eyes find it challenging to observe tissue motions beneath the skin and to follow these motions in real-time manually, ultrasound technology can detect deep tissue motions, and robotic systems can move probes swiftly with high precision. The integration of ultrasound and robotics offers promise for tracking respiratory motion of deep tissues for ULM, which is within the research field known as robotic ultrasound (RUS) \cite{von2021medical, jiang2023robotic}. Existing RUS studies address various applications, including optimising image quality through probe orientation \cite{chatelain2017confidence, welleweerd2021out}, generating scan paths to reconstruct large organ views \cite{huang2021towards, jiang2024needle}, adjusting probe force for elastography \cite{schneider2012remote, yao2019three}, and stabilising imaging and object tracking \cite{nadeau2014intensity, royer2017real}. Robotic ultrasound for managing respiration motion in ULM imaging is an object tracking problem. Most existing studies on object or target tracking have been demonstrated through offline simulations \cite{nadeau2014intensity, royer2017real}, using low-frame-rate commercial machines (e.g., 25 Hz). A few studies have achieved real-time probe targeting during imaging, with frame rates around 30 Hz \cite{chatelain2017confidence} and volume rates of 13 Hz \cite{nadeau2013intensity} or 2 Hz \cite{nadeau2016moments}. These studies relied on ensuring that the sampling period of the control action—including motion detection and robot control—was shorter than the time interval between frames. 

%(Challenges in using robotic ultrasound for dealing with respiration in ULM) 

However, ultrafast ultrasound, beneficial to ULM, typically operates at much higher frame rates compared to low-frame-rate commercial machines, resulting in significantly shorter time intervals between frames. The existing control strategies for robotic systems are rendered inadequate, as they fail to complete tasks within the critical window of two frames. This issue is further exacerbated by the use of 4D (3D + time) ultrasound devices, which can capture motions out of 2D imaging plane but generate an overwhelming amount of data that current control systems struggle to process efficiently. Moreover, the time required for volumetric beamforming is substantial, often exceeding the frame interval in ultrafast ultrasound by several times. Thus, developing new control strategies capable of real-time high-frame-rate acquisition, data processing, and robotic control is imperative. These advancements are crucial to making breath-holding unnecessary for ULM imaging with robotic systems.

%(Summary of our work) 

This study proposes a new approach for enabling ULM in the presence of large tissue motion, by utilising high-frame-rate volumetric ultrasound combined with robot-assisted target tracking in an asynchronous manner. The strategy was validated by reconstructing 3D SR images from a microvasculature phantom moved by motorised translation stages over a range of 20 mm—twice the aperture size of the matrix array mounted on a robot arm. %The phantom was  Image processing tasks, including volumetric image reconstruction and motion detection, were accelerated using GPU, achieving real-time performance with processing times ranging from 0.08 to 0.35 seconds. Real-time image processing and robot control were performed asynchronously with ultrasound data acquisition, maintaining a constant frame rate of 85 Hz. After real-time tracking, residual motion in the image view was reduced to 1.12 ± 0.67 mm for a target moving at 5 mm/s. Post-processing of the acquired data—including correction for residual motions, clutter filtering, MB localisation, and tracking—further demonstrated the feasibility of the proposed strategy for ULM imaging.

\section{Methods}
\begin{figure*}
    \centering
   \includegraphics[width=16cm]{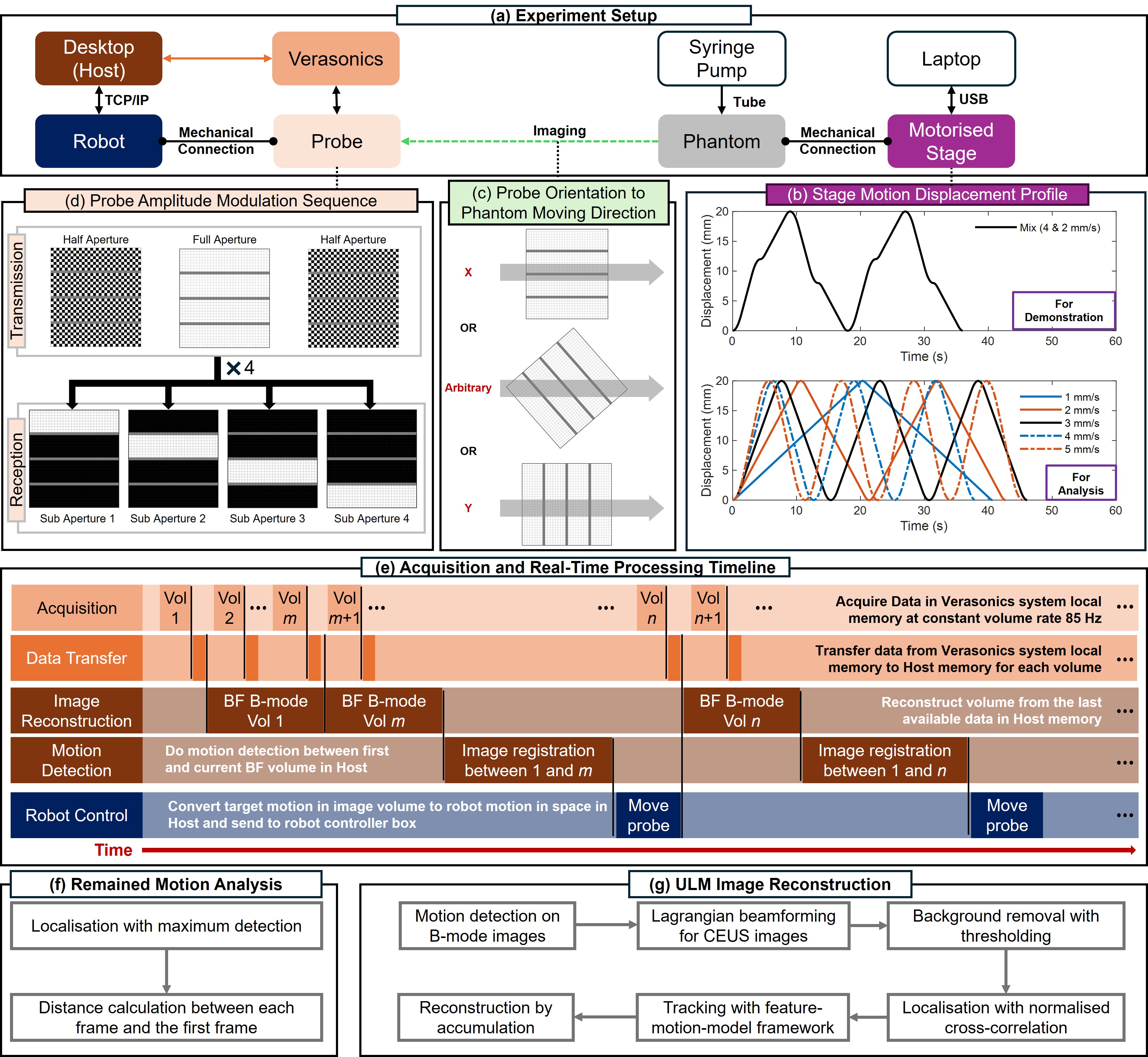}
    \caption{Study diagram. (a) Experiment setup: a matrix array was fixed on the tool base of a robot arm by a 3D printed holder; the robot and the Verasonics system were connected to the same desktop (host); the robot and the host were communicated via TCP/IP; a phantom with crossed tubes was placed on motorised translation stages that were controlled by a laptop; MBs were injected into the phantom by a syringe pump. (b) Phantom was moved in six different motion profiles. (c) Phantom motions were given in three different directions along the matrix array. (d) Half-Full-Half aperture transmission and sub-aperture reception were used for the multiplexed matrix array. \textcolor{black}{(e)} Designed asynchronous real-time motion tracking timeline, consisting of data acquisition, image processing and robot control. (f) Motions remained in images after real-time motion tracking were analysed to test the motion tracking performance for different target motion speed levels. (g) The feasibility of ULM reconstruction was verified when target motion was double the probe aperture size.}
    \label{fig:Diagram}
\end{figure*}

This study consists of three technical components: 1) 4D ultrasound data acquisition for ULM with robot-assisted motion tracking in real time; 2) analysis of residual motion in imaging view after real-time motion tracking; 3) ULM image reconstruction from the data acquired with robotic ultrasound. Diagram of this study is shown in Fig. \ref{fig:Diagram}.

This section begins with an introduction to the experimental setup, followed by detailed descriptions of data acquisition with real-time robot-assisted motion tracking, residual motion evaluation after real-time tracking, and ULM image reconstruction under double-aperture-size motion conditions.

\subsection{Experiment Setup}

\subsubsection{Phantom preparation}
A cross-channel phantom was made by wire-templating method to mimic microvasculature in tissue \cite{kawara2023capillary}. Pulled and tapered glass capillaries (B100-50-10, World Precision Instruments, UK) were coated with 3-(Trimethoxysilyl) propyl methacrylate (TMSPMA, Sigma-Aldrich, UK), used to guide the inlets and outlets. Nichrome wire with a diameter of 180 µm was vapour-deposited with Parylene C (SCS, USA) at 10 µm using a vapour deposition system (PDS 2010 Parylene Deposition System, SCS, USA). An alignment scaffold for supporting the glass capillary and hydrogel was 3D printed using the SLA resin (Clear Resin V4, FormLabs, USA) with a printer (FormLabs 3+, USA). Glass capillary grips were moulded with polydimethylsiloxane (PDMS) (SYLGARD™ 184 Silicone Elastomer Kit, Dow Corning, US) in the reservoirs in the alignment scaffold. The Parylene-coated template wires were then threaded through the capillaries held by the moulded PDMS. 5\% w/v  Acrylamide and 0.3\% w/v  bis-acrylamide (Sigma-Aldrich, UK) were dissolved in deionised water.  This solution was then mixed with 0.1\% w/v  of ammonium persulfate (Sigma-Aldrich, UK) and 0.1\% v/v  of tetramethyl ethylenediamine (TEMED) (Sigma-Aldrich, UK) and poured into the alignment scaffold in the reservoir and left for 1 hour at room temperature to polymerise. Wires were slowly removed from the polymerised gel to form the wall-less cross-channel phantom.
A cross-channel phantom was made by wire-templating method to mimic microvasculature in tissue \cite{kawara2023capillary}. Pulled and tapered glass capillaries (B100-50-10, World Precision Instruments, UK) were coated with 3-(Trimethoxysilyl) propyl methacrylate (TMSPMA, Sigma-Aldrich, UK), used to guide the inlets and outlets. Nichrome wire with a diameter of 180 µm was vapour-deposited with Parylene C (SCS, USA) at 10 µm using a vapour deposition system (PDS 2010 Parylene Deposition System, SCS, USA). An alignment scaffold for supporting the glass capillary and hydrogel was 3D printed using the SLA resin (Clear Resin V4, FormLabs, USA) with a printer (FormLabs 3+, USA). Glass capillary grips were moulded with polydimethylsiloxane (PDMS) (SYLGARD™ 184 Silicone Elastomer Kit, Dow Corning, US) in the reservoirs in the alignment scaffold. The Parylene-coated template wires were then threaded through the capillaries held by the moulded PDMS. 5\% w/v  Acrylamide and 0.3\% w/v  bis-acrylamide (Sigma-Aldrich, UK) were dissolved in deionised water.  This solution was then mixed with 0.1\% w/v  of ammonium persulfate (Sigma-Aldrich, UK) and 0.1\% v/v  of tetramethyl ethylenediamine (TEMED) (Sigma-Aldrich, UK) and poured into the alignment scaffold in the reservoir and left for 1 hour at room temperature to polymerise. Wires were slowly removed from the polymerised gel to form the wall-less cross-channel phantom. \textcolor{black}{The attenuation coefficient of the phantom is about 0.17 dB/(MHz*cm) \cite{cafarelli2017tuning}.}

A ball of 1 mm in diameter and its supporting structure were 3D printed in the alignment scaffold to provide a landmark useful for online motion tracking and offline motion analysis. Home-made MBs \cite{li2015quantifying} were diluted to a concentration of $\sim$ 1 ×  10$^7$  bubbles/ml and pumped into the channels by a syringe pump  (Harvard Apparatus, USA) \textcolor{black}{at a flow rate of 1.5 µL/min} to mimic blood flow and verify if data acquired from a moving target was feasible for ULM reconstruction. 

\subsubsection{Hardware connection}
\begin{figure}
    \centering
   \includegraphics[width=8.5cm]{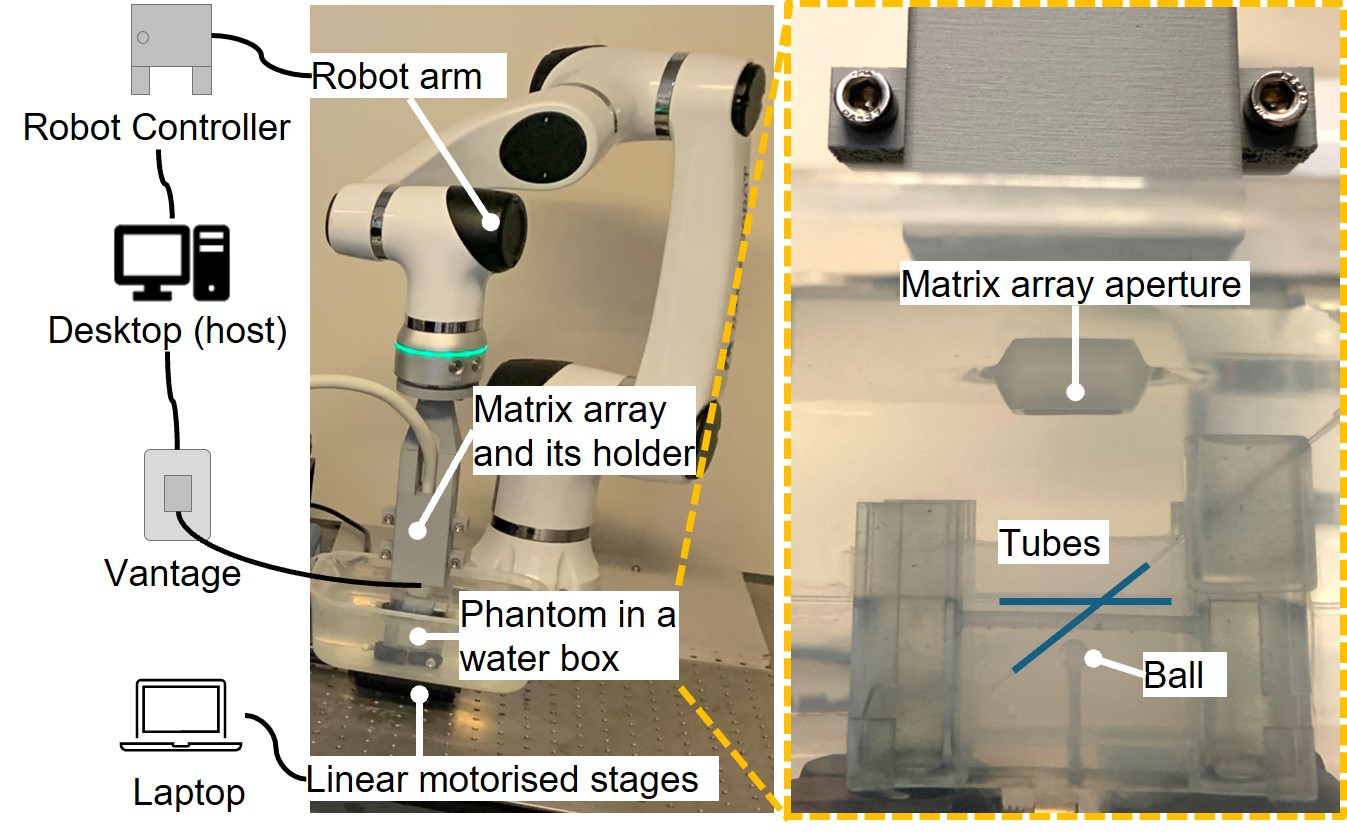}
    \caption{Experimental Setup, diagram of which is shown in Fig. \ref{fig:Diagram}(a). The left side shows the overall setup. The right side shows the aperture of ultrasound array and the home-made microvasculature phantom.}
    \label{fig:setup}
\end{figure}

A matrix array (8MHz, 32 × 32 elements, 9.6 mm × 10.6 mm aperture, Vermon, France) was used to achieve 4D  ultrasound imaging on the microvasculature phantom, as shown in Fig. \ref{fig:setup}. The 1024-channel matrix array was connected to a 256-channel ultrasound research system (Vantage 256, Verasonics, USA) via 4-in-1 multiplexing. The probe was fixed on the tool base of a robot arm (Elfin05, Han's Robot, China). The phantom was placed on two motorised translation stages (PT1-Z9, Thorlabs, USA). Each of the stage can achieve a maximum speed of 2.6 mm/s, and one was stacked on top of the other and they were programmed to move at the same time. 

The Vantage system was connected with the host computer as provided by Verasonics. The control box of the robot arm was connected to the host via an Ethernet cable. The motorised stages were connected neither to the Vantage system nor to the host but to a separate laptop. The data acquisition was manually started by a designed Vantage interface on the host, and then motion of the motorised stage was manually started by running Kinesis software (Thorlabs) on the laptop.

\subsubsection{Motion protocol}
 The phantom was moved by the stages in a range of 20 mm which was a reported liver respiration motion range \cite{fahmi2018respiratory} and was double the aperture size of the matrix array. The phantom was moved at different speeds with a fixed acceleration and deceleration, 3 mm/s$^2$, as shown in Fig. 1 (b). 
\begin{enumerate}
    \item To demonstrate the target lost in the imaging FoV under large motions and the feasibility of keeping target in the FoV with robot-assisted motion tracking, the phantom was moved at varying speed levels, between 4 mm/s and 2 mm/s, within each round trip.
    \item To test the performance of real-time target tracking with 4D ultrasound, the phantom was moved in five round trips at a speed from 1 to 5 mm/s with a step increase of 1 mm/s in each round trip. The performance-testing trials were repeated. 
\end{enumerate}	

\textcolor{black}{Target moving along axial direction can be kept in FoV with enough imaging depth and signal-to-noise ratio that can be electrically adjusted.} Compared to targets moving along the axial direction, targets moving in the elevation and lateral directions relative to the probe are more likely to exit the imaging FoV, \textcolor{black}{because elevation and lateral dimensions are physically restricted by aperture size and element directivity and are generally smaller than imaging depth. To verify if the target tracking is feasible to keep a target in FoV when the targets move at multiple directions}, the target moving direction was aligned along the lateral, elevational or arbitrary direction of the matrix array by rotating the probe around its depth axis, as shown in Fig. \ref{fig:Diagram} (c), to verify if the target tracking is feasible for multiple directions.

\subsection{Data Acquisition with Real-Time Robot-Assisted Motion Tracking}

\subsubsection{Data acquisition}
To reconstruct B-mode and CEUS image volumes, Amplitude Modulation (AM) was implemented with Half-Full-Half \textcolor{black}{aperture} transmission sequence. \textcolor{black}{The full-aperture transmission was achieved by using all the 1024} \textcolor{black}{elements, and the half-aperture transmission was achieved by using two different halves of elements, i.e., 512 elements, interleaved in the matrix array}, as shown in Fig. \ref{fig:Diagram} (d). Only a single non-steering plane wave was transmitted from each aperture at a centre frequency of 7.8 MHz, to minimise the number of transmissions and data storage. Four multiplexed sub-apertures (32 × 8 elements) received echoes in sequence after the three AM pulses finished for each receiving aperture, to minimise influence of phase shifts due to motions on tissue signal cancellation. The time interval between pulses for reconstructing one volume was empirically set as four times the time of flight for an imaging depth of 45 mm, to avoid reverberations. Duration of each acquisition was set to 70 s to cover the target motion shown in Fig. \ref{fig:Diagram} (b). The acquisition volume rate was set to 85 Hz, which could be higher but was empirically selected as such a value to prevent acquired data being too large, resulting in about 69 Gb of data being saved in one 70-second acquisition. 

\subsubsection{Online data processing}
4D ultrasound images were reconstructed in large voxel size and small volume, and motion detection was simplified to achieve real-time data processing. B-mode volumes were reconstructed with a voxel size of two wavelengths, i.e., 0.4 mm × 0.4 mm × 0.4 mm, by Delay and Sum (DAS) beamformer we developed in CUDA (Toolkit Version 11.8, compiled by ptx in MATLAB R2023a). Considering that probe motion is rigid and target after rotation is still in the imaging view, only translations of the target, i.e., three Degrees of Freedom (DoF), were detected by image registration (MATLAB function \textit{imregtform} with a regular step gradient descent and three pyramid levels) to minimise number of parameters in optimisation and save computation cost. A large image volume (45 mm × 20 mm × 20 mm) was first reconstructed to view the target. \textcolor{black}{The part of volume exceeding the footprint was reconstructed, even if the intensity and resolution in this part was lower than those beneath the footprint, for the convenience of locating the target.} Then, a small cuboid image volume, \textcolor{black}{around 12.5 mm × 12.5 mm × 12.5 mm and} and containing about 3 × 10$^4$ voxels, was manually cropped around high-intensity landmarks for real-time reconstruction and motion detection, to save the computation cost during the real-time tracking. 

\subsubsection{Robot motion control}
In this paper, a simple and easy-to-implement robot control strategy was used to evaluate the feasibility of using high-frame-rate volumetric ultrasound combined with robot-assisted target tracking. The robot was controlled to adjust the probe to maintain zero translation of the target within the imaging FoV. Target motion was detected between the initial and each subsequent reconstructed volume, and this data was used to update the spatial coordinates in the robot's workspace, thereby repositioning centre of the imaging FoV. To effectively track the target's motion, the robot's speed needed to exceed that of the target. Considering the robot's inertia and the stiffness of the 3D-printed probe holder, the robot's movement was empirically set with an acceleration of 250 mm/s² to a maximum speed of 50 mm/s, followed by a deceleration of -250 mm/s² to a stop, preventing probe oscillation. With these settings, the minimum distance required for the robot to reach maximum speed was approximately 10 mm, longer than the distance the probe needed to move for each experimental iteration. Consequently, the probe exhibited only acceleration and deceleration, without reaching steady state motion, as the robot repositioned the centre of the FoV. The robot was controlled using a self-developed MATLAB interface, with communication between the host computer and the robot control box managed via TCP/IP.

\subsubsection{Data acquisition and processing timeline}

Using a host computer equipped with a 16-core CPU@4GHz (Ryzen 5955WX, AMD) and a 2560-CUDA-core GPU@1.78GHz (GeForce RTX 3060, NVIDIA) for one acquisition, beamforming required about 0.04 seconds. Motion detection took between 0.04 and 0.3 seconds, depending on the convergence of iterations in the image registration process, while robot control varied from 0.03 to 0.25 seconds based on the movement distance. The total time for one round of data processing varied, with a minimum of 0.11 seconds, which is about 10 times the time interval between frames at the desired volume rate of 85 Hz. To ensure data acquisition at a constant frame rate— essential for accurate flow speed measurement in ULM processing— data acquisition and real-time data processing were executed asynchronously, as illustrated in Fig. \ref{fig:Diagram}(e). Beamforming was performed on the latest available frame in the host memory. When both the initial frame and subsequent frames were available, image registration was carried out between the first and following frames, and the robot executed the generated motion commands. This process of beamforming, motion detection, and robot control was repeated after each round of processing for the entire acquisition.

\subsubsection{Safety design}
In a laboratory setting where the robot operates while humans are present, safety is crucial. The safety design for the robot-assisted tracking was implemented at four levels: high-level movement command constraints, low-level motor control constraints, an emergency stop button, and a remote control socket. The robot's safe workspace was defined as [-50 mm, 50 mm] in the lateral and elevational directions and [-5 mm, 5 mm] in the depth direction relative to the probe. On the host computer, the centre of this safe workspace was set manually through a button in the designed interface and was also automatically established at the start of each data acquisition. Movement commands were not sent to the robot if the calculated spatial coordinates fell outside the defined safe workspace. Additionally, the safe workspace parameters were communicated to the robot control box to enforce these restrictions at the low level. An emergency stop button, provided by the manufacturer, could cut off power to the robot arm's motors, while a remote control socket could disconnect power to the entire robot.

\subsection{Residual Motion Evaluation after Real-Time Tracking}

Motions of the imaging target in the high-frame-rate 4D ultrasound images could not be completely eliminated by the real-time robot-assisted tracking, due to the control delay and speed difference between the motions of the target and the probe. To evaluate the remaining motion, B-mode volumes were reconstructed with a voxel size of 0.1 mm × 0.1 mm × 0.1 mm, and the landmark scatter in the phantom was localised by weighted centroiding. The displacement between the scatter’s lateral and elevational position in each frame and the first frame was calculated for each trial. The onset of stage movement in each acquisition was detected when the scatter moved in the FoV over 0.1 mm.  The mean and standard deviation of the displacements only occurring during stage movement were calculated.

 \begin{figure}
    \centering
   \includegraphics[width=8.5cm]{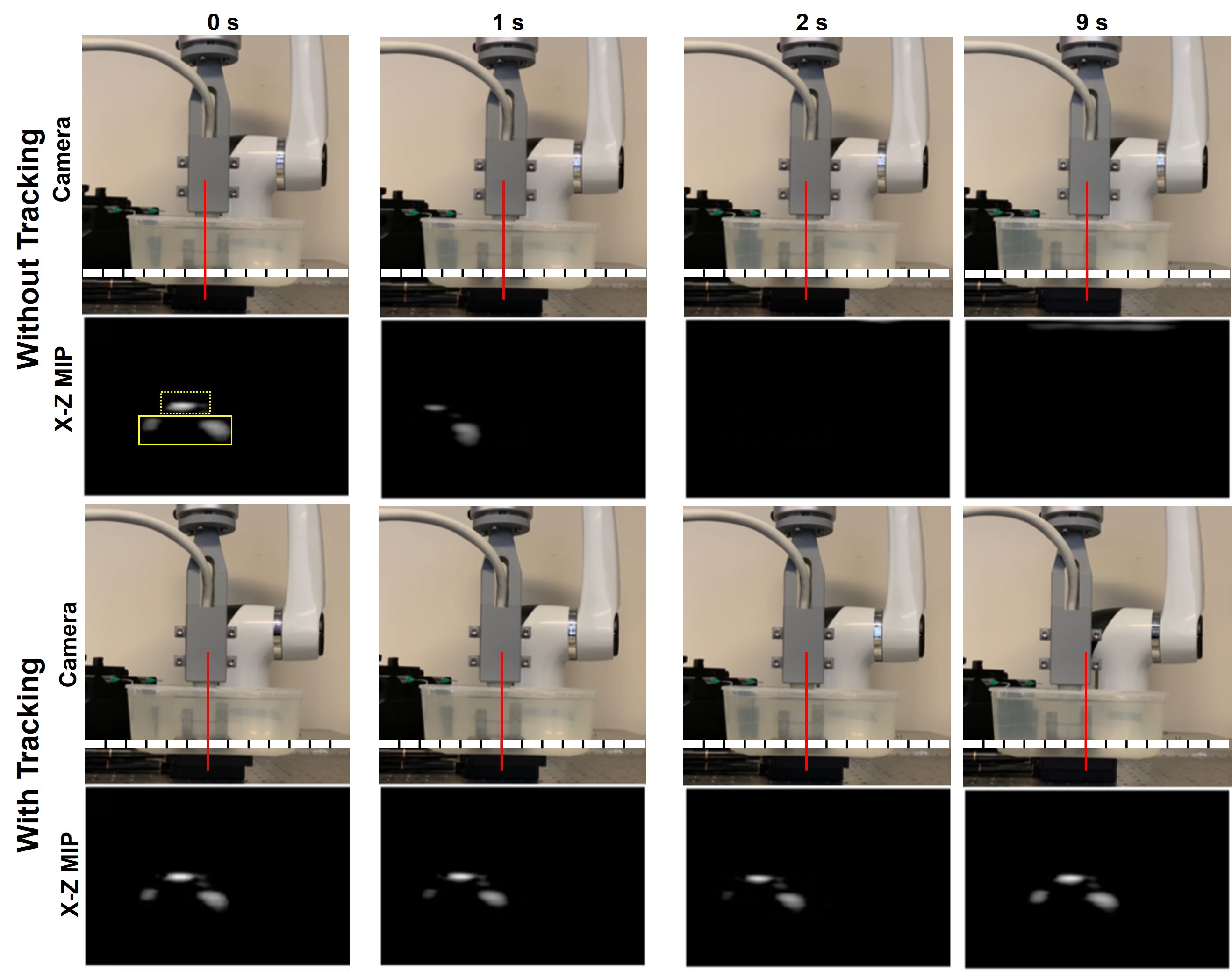}
    \caption{Camera videos and corresponding lateral-depth (X-Z) maximum intensity projection (MIP) of reconstructed 4D ultrasound images with or without robotic tracking. The object in the dashed yellow box of the MIP ultrasound image is the landmark ball scatter in the phantom and the lobes in the solid yellow box are the structure supporting the ball. This landmark moved out of the imaging FoV when the probe is fixed but kept in the FoV when the robotic tracking is on. The imaging FoV is centred on the array central axis with a lateral length of 15 mm. Log compression and a dynamic range of 20 dB was used in the visualisation. The red line denotes a reference fixed in the camera view of each row.}
    \label{fig:screen_shots}
\end{figure}
 \begin{figure*}
    \centering
   \includegraphics[width=17cm]{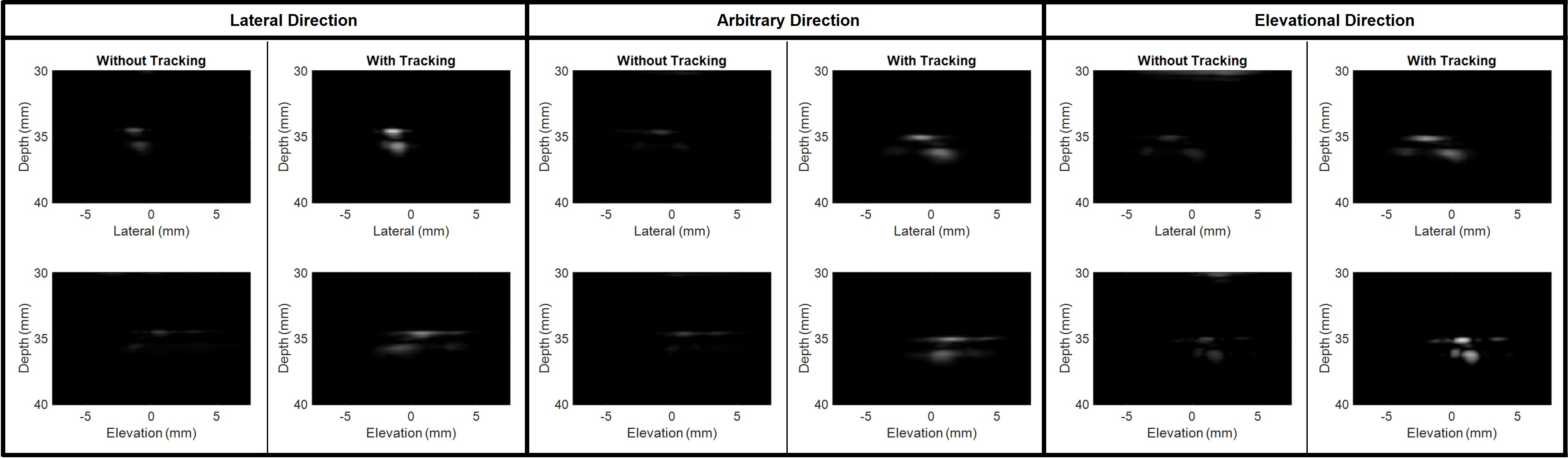}
    \caption{Temporal average of reconstructed B-mode volume \textcolor{black}{ Maximum Intensity Projection (MIP)} of the landmark to demonstrate difference with robotic tracking or not. The images with the robot tracking are brighter than those without the robot tracking, meaning more frames were overlapped with the robot tracking.}
    \label{fig:B_mode_demo}
\end{figure*}
\begin{figure*}
    \centering
   \includegraphics[width=16cm]{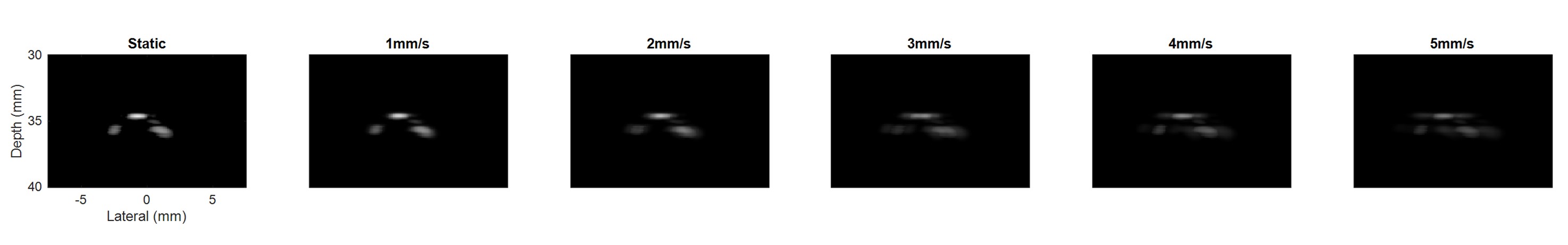}
    \caption{Temporal average of reconstructed B-mode volume MIP of the landmark when moving target along probe’s lateral direction at different speeds. Images were less overlapped among frames when target moved faster.}
    \label{fig:B_mode_testing}
\end{figure*}

\subsection{ULM Image Reconstruction with Double-Aperture-Size Motion}
\subsubsection{Residual motion correction}
Data acquired during the motion with varying speed levels were processed to demonstrate the feasibility of ULM imaging under large motions. The residual motion was detected and corrected for reconstructing ULM images. Data acquired with full aperture transmission was reconstructed for B-mode images by the above-mentioned DAS beamformer with a voxel size of 0.1 mm × 0.1 mm × 0.1 mm.  Motion detection was done between the first frame, i.e., the reference frame, and each of following frames by self-developed 3D rigid image registration codes implemented in CUDA, whose cost function was the sum of square intensity difference and solver was the Levenberg–Marquardt (LM) algorithm. The converged output of the LM algorithm was set as the initial for the LM iterations when processing the next frame, which helped improve the convergence and reduce the computation cost of iterations to find minimum.

\subsubsection{CEUS image reconstruction}
MB signals were obtained by AM modulation, which was done through subtracting channel signals acquired with full transmission aperture by the sum of channel signals acquired with the two half apertures. MB images were reconstructed by our previously proposed Covariance to Variance Beamformer\cite{yan2023fast} with a voxel size of 0.1 mm × 0.1 mm × 0.1 mm. Considering doing motion correction on reconstructed CEUS images might generate artefacts during interpolating low-resolution pixels, tissue motions were corrected by reconstructing the CEUS images in Lagrangian coordinates, where image values were reconstructed with their physical positions but assigned to the corresponding voxels in the reference frame. 

\subsubsection{MB localisation and tracking}
Background noise and side lobes were reduced by assigning zero to voxel whose intensity was below either an adaptive threshold, i.e., weighted and averaged surrounding voxel intensity, or an empirically given noise level. Imaging PSF was estimated by averaging five manually segmented and self-normalised single MB images. Normalised cross-correlation (NCC) was done between each volume and the estimated PSF. MBs were localised by detecting peaks over 0.5 on the NCC coefficient map, interpolating voxels surrounding each peak by 5 times via cubic interpolation and detecting maximum position in each interpolated region. MBs between frames were paired by our previously proposed feature-motion-model tracking framework \cite{yan2022super,yan2024transthoracic}. A persistence filter with a threshold of 4 frames was used to further reduce noise. ULM images were reconstructed by accumulating MB trajectory number for density map and averaging MB trajectory speed for speed map with a voxel size of 0.02 mm × 0.02 mm × 0.02 mm. For visualisation, the localisation density map was smoothed by a 3D Gaussian whose Full Width of Half Maximum was 40 µm, and the speed map was smoothed by a 3D ball whose diameter was 60 µm.

\section{Results}
The robot-assisted motion tracking demonstrated its effectiveness in keeping the moving target in the imaging FoV, as shown in Fig. \ref{fig:screen_shots} and the supplementary video. While keeping the probe static as commonly used for ULM acquisition, the landmark structures including the ball and the supporting structure in the phantom were in the first frame but went out of the imaging FoV in following frames. With the robot-assisted motion tracking, the landmarks were always within the imaging view. Demonstrations for the whole stage motions along all the three directions can be found in the supplementary video.

B-mode volumes were log-compressed with the maximum intensity in the first frame, rescaled to grey scale with a dynamic range of 20 dB, spatially projected to the lateral-depth (X-Z) view and elevation-depth (Y-Z) view by the maximum intensity and temporally averaged across frames, as shown in Fig. \ref{fig:B_mode_demo} and \ref{fig:B_mode_testing}. \textcolor{black}{When the target motion duration in images } \textcolor{black}{  varied across different stage motion profiles, frames without significant target motion were excluded from the averaging for the visualized demonstration. For Fig. \ref{fig:B_mode_demo}, 3400 frames from each acquisition were used for averaging. For Fig. \ref{fig:B_mode_testing}, the number of frames used for averaging at speed levels of 1, 2, 3, 4, and 5 mm/s were 4250, 4450, 4750, 3950, and 4850, respectively.}  Fig. \ref{fig:B_mode_demo}  shows the temporally average across the phantom moving with varying speed levels, i.e. frames in the supplementary video. The temporally averaged images with robotic tracking are much brighter than those without the robotic tracking, indicating more frames containing the landmark are overlapping when robotic tracking is on, which is in agreement with Fig. \ref{fig:screen_shots} and the supplementary video.

\begin{figure}
    \centering
   \includegraphics[width=8.5cm]{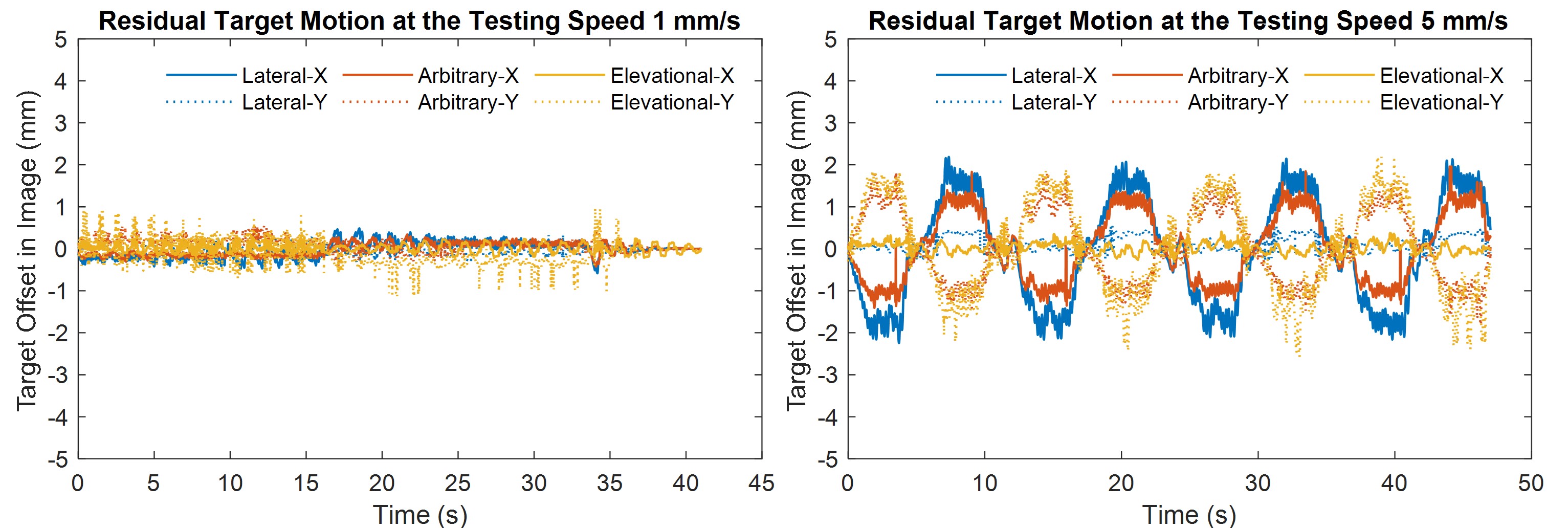}
    \caption{Target motions in the imaging view after real-time motion tracking.}
    \label{fig:residual_motion_curve}
\end{figure}

Fig. \ref{fig:B_mode_testing} shows the results when the phantom moved at different speed levels. The temporally averaged image becomes more spread and darker when phantom moving speed level increased. Fig. \ref{fig:residual_motion_curve} presents the landmark position shifts to its initial position in imaging FoV when the phantom moved at the lowest and the highest speed levels respectively. It can be seen that most of motions remained after real-time tracking are within 2 mm.

\begin{figure}
    \centering
   \includegraphics[width=8.5cm]{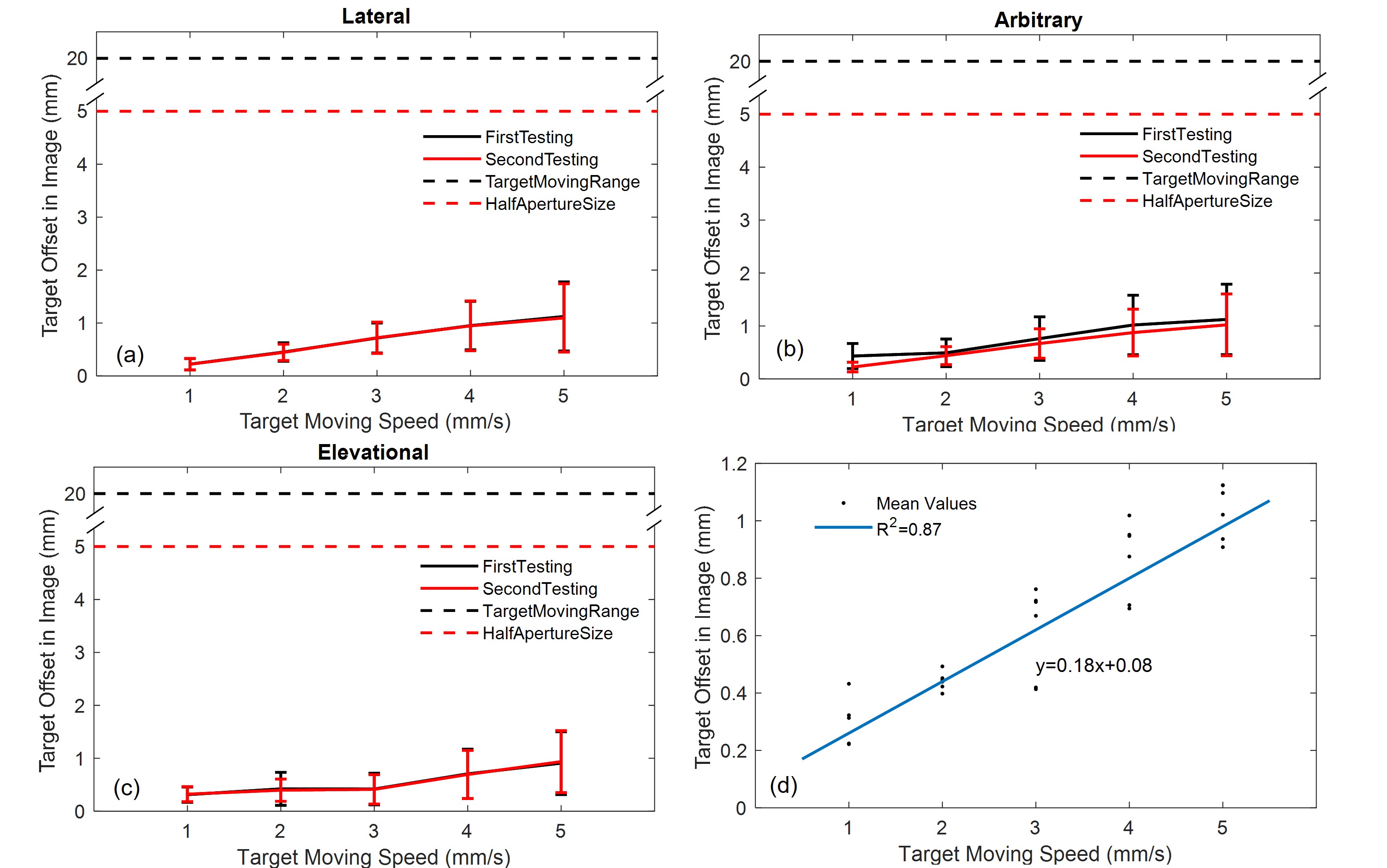}
    \caption{Residual motions in the imaging view when tested at different moving speeds and directions (a-c), and linear regression between the mean values of motions and the tested moving speeds (d). The red dashed line denotes the size of half aperture. and the black dashed line denotes the maximum distance that the phantom was moved to.}
    \label{fig:residual_motion_mean}
\end{figure}

Fig. \ref{fig:residual_motion_mean} (a-c) show the mean and standard deviations of the residual moving distance after excluding the data where the phantom did not move. The mean residual motions increase with the phantom moving speed. The residual motions at the highest speed level were 1.12 ± 0.65 mm and 1.10 ± 0.65 mm for lateral direction, 1.12 ± 0.67 mm and 1.02 ± 0.58 mm for the arbitrary direction, and 0.90 ± 0.59 mm and 0.93 ± 0.56 mm for the elevational direction, around 20\% the half aperture size. The mean of residual motions versus different speed levels were regressed by a linear function shown in Fig. \ref{fig:residual_motion_mean} (d). 

\begin{figure}
    \centering
   \includegraphics[width=15cm]{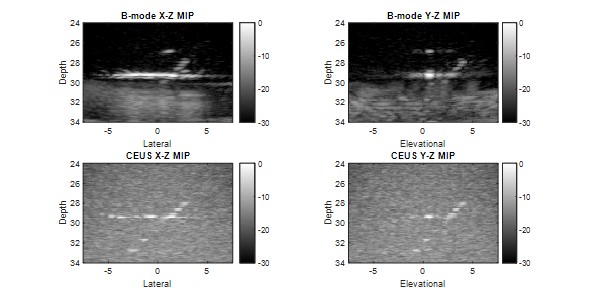}
    \caption{B-mode and CEUS images of tube signals.}
    \label{fig:AM_demo}
\end{figure}

\begin{figure*}
    \centering
   \includegraphics[width=14.5cm]{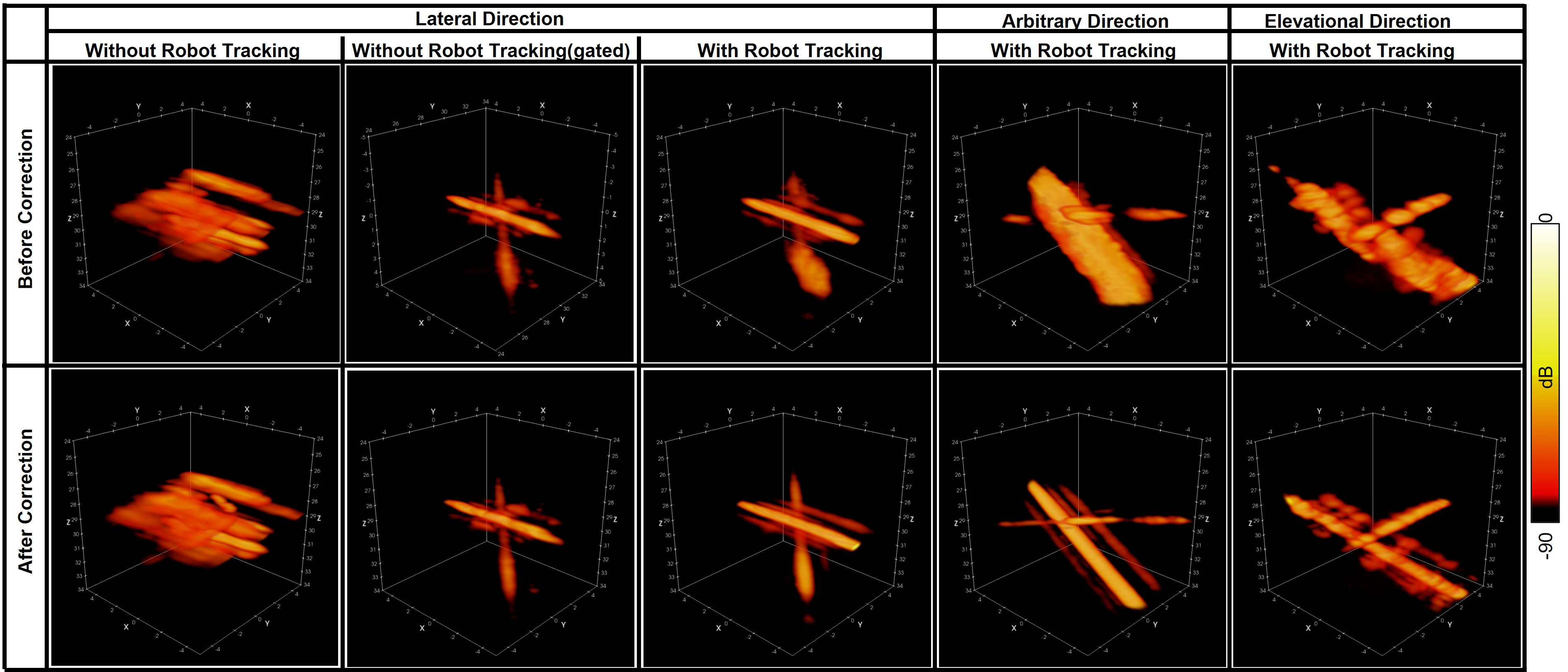}
    \caption{\textcolor{black}{Power doppler images to demonstrate the effectiveness of robot-assisted motion tracking. The phantom was moved by the Mix displacement profile shown in Fig. \ref{fig:Diagram} (d)} }
    \label{fig:power_doppler_images}
\end{figure*}

\begin{figure*}
    \centering
   \includegraphics[width=14.5cm]{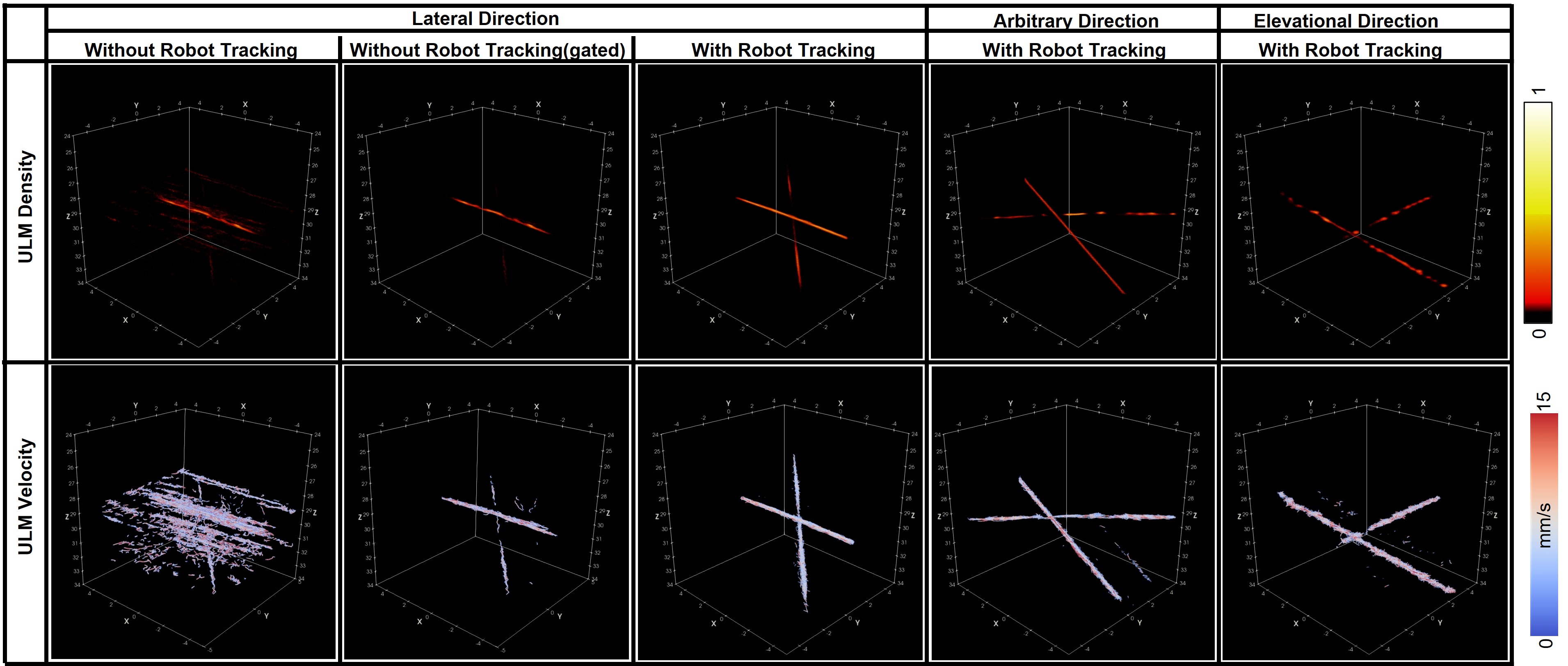}
    \caption{\textcolor{black}{ULM images to demonstrate feasibility of achieving super-resolution imaging under large motions with the developed technique. ULM images of one data acquired without robot tracking was also reconstructed for comparison. The phantom was moved by the Mix displacement profile shown in Fig. \ref{fig:Diagram} (d)} }
    \label{fig:ULM_images}
\end{figure*}

\begin{figure}
    \centering
   \includegraphics[width=7.5cm]{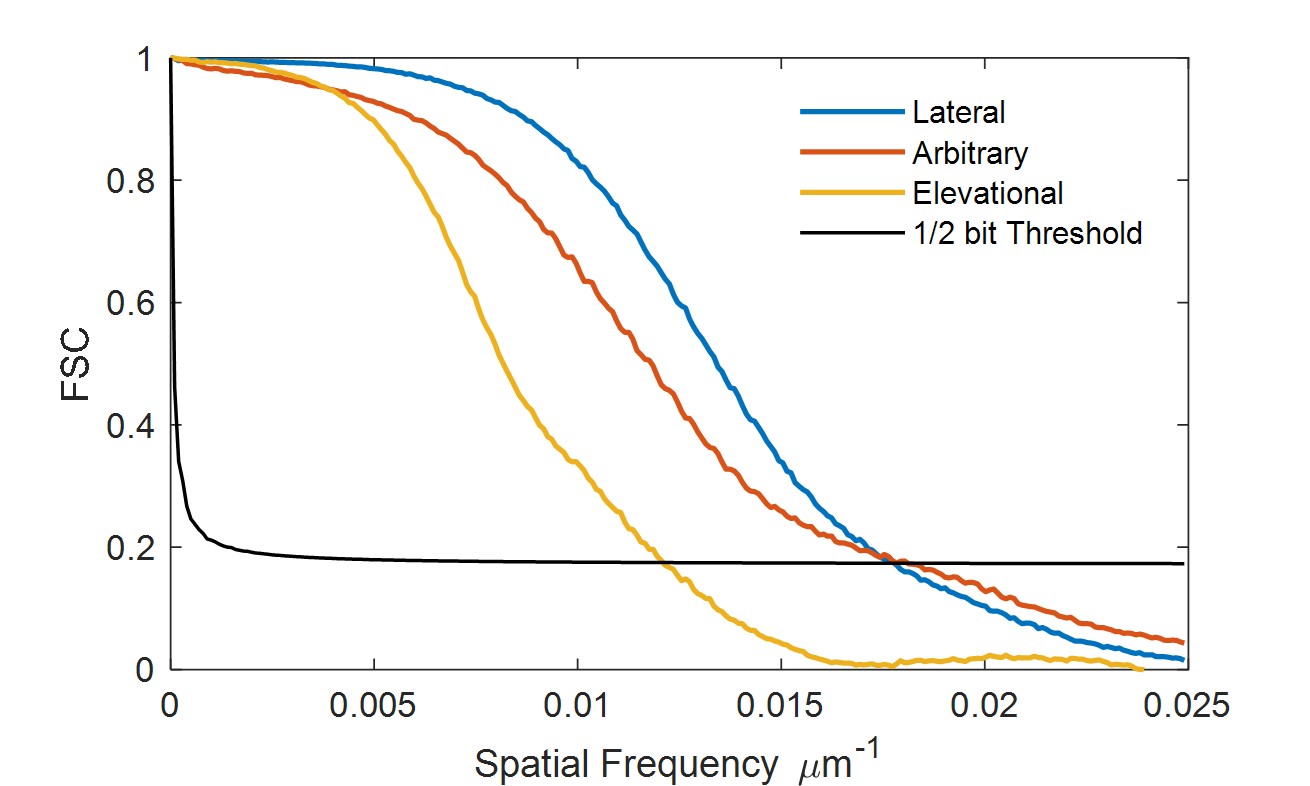}
    \caption{FSC analysis of SR images obtained at three different directions.}
    \label{fig:FSC}
\end{figure}

\textcolor{black}{Fig.\ref{fig:AM_demo} show the ultrasound images of the phantom before and after AM processing. Power Doppler and ULM images were reconstructed from data acquired with robot tracking, as shown in Fig. \ref{fig:power_doppler_images} and \ref{fig:ULM_images}.  Power Doppler and ULM images of one data without robot tracking were also reconstructed for comparison. For the data without robot tracking, it is difficult to see crossed-tube structure in the both the Power Doppler and ULM images, as not only tubes moved out of the FoV but also strong reflecting signals from the phantom wall moved into the FoV.  After gating frames with the ball centre within the FoV, crossed-tube structure can be clearly seen. However, the gated data only contains 2550 static frames and 900 moving frames, almost half of the data is excluded from accumulation, so the image must be reconstructed with less saturation than the complete data. Power Doppler images of all the frames in three datasets acquired with robot tracking presents clear } \textcolor{black}{  tube structures. Even if each of these three datasets contains more moving frames than the gated one,} motion correction in the post-processing could concentrate main lobe signals along two straight lines, \textcolor{black}{as shown in the second row of Fig. \ref{fig:power_doppler_images}}.
It is worth noting that the weak signals parallel to the strong signals were side lobes due to the uneven distributed elements in footprint of the matrix array, as shown in Fig. \ref{fig:Diagram} (c) and (d). The discontinuity of crossed channel signals might result from that channels were across or at the low-pressure regions under the non-active rows in the footprint. The SR density and speed maps reconstructed from MB localisation and tracking were also shown in Fig. \ref{fig:ULM_images}. The reconstructed speed maps visually present more continuous structure and noise than the density, as the two maps were reconstructed via averaging and accumulating respectively. Fourier Shell Correlation (FSC) analysis was done between two SR density maps reconstructed with odd and even number of trajectories \cite{hingot2021measuring}. The FSC resolutions measured at the 1/2-bit threshold were 56.5, 55.6, and 83.3 µm for three SR density maps in order, as shown in Fig. \ref{fig:FSC}. It is demonstrated that super-resolution imaging could be achieved by the proposed method even if the target moved in a range double the probe aperture size.

\section{Discussion}

%(Overall Contribution) 

This work is the first feasibility study to demonstrate the use of a robot for tracking large target motions during high-frame-rate data acquisition for ULM. For the first time, a robot-assisted 4D ultrasound data acquisition framework and system has been developed that relies solely on ultrasound imaging modalities, without incorporating additional imaging techniques. The system required only an Ethernet cable to connect the commercial robot arm to the ultrasound research system. Importantly, the robot arm is significantly more affordable than the matrix array, providing a cost-effective solution for the ultrasound community. 

%(Technical Contributions to deal with challenges) 
\textcolor{black}{While the motion at the body surface can be correlated with the motion of the imaging target, their movements are not always the same. As highlighted in the literature\cite{fahmi2018respiratory}, the imaging target's motion cannot be directly or reliably estimated } \textcolor{black}{ from markers on the body surface. Various techniques have been developed to predict target motion using trained models derived from collected data \cite{mcclelland2013respiratory}; however, their accuracy often depends on the similarity between the training data and the prediction scenarios. In contrast, robotic tracking, which directly detects deep target motions using ultrasound, provides greater precision. This method is particularly advantageous in scenarios involving sliding shifts between the skin and underlying organs, where surface-based estimations become unreliable. The traditional image registration generally corrects motion by image interpolation after acquisition, but target moving out of FoV cannot be recovered by the interpolation; the proposed approach corrects motion by moving with target physically during acquisition, and thus information loss can be reduced. While FoV of probe can also be enlarged with diverging waves to cover the target motion, plane wave is less affected by element directivity and has higher resolution and SNR than diverging waves, especially at deep regions, which can benefit the microbubble isolation and localisation. Therefore, the proposed method is used with plane wave to keep target in FoV and can also be used with diverging waves if required. Tracking target motion can also help reduce difficulty in offline motion correction for ULM imaging.}

The main challenge in robot-assisted motion tracking for ULM is achieving real-time motion tracking while maintaining a constant frame rate during data acquisition. In this study, this challenge is addressed through an asynchronous data acquisition and processing strategy, as illustrated in Fig. \ref{fig:Diagram}(e). This approach allows high-frame-rate acquisition regardless of the data processing speed. Given that longer data processing times can lead to greater target motion in the imaging FoV, several engineering practices were implemented to enhance data processing speed and achieve real-time performance.

To minimise the complexity and computational demands of motion detection, only the motions crucial for maintaining the imaging FoV on the target—specifically translations—were estimated. While rigid image registration typically requires more processing time compared to optical flow and Doppler-based methods for the same data volume, it is less sensitive to variations in image resolution, motion scale, and time intervals between beamformed frames. Real-time volumetric beamforming was achieved by using a large voxel size (2 wavelengths in this study), and the processing speed was further improved by reducing the reconstructed volume to only include high-intensity signals for image registration. The volumetric image reconstruction was accelerated using GPU-based processing with a self-developed beamformer, and real-time motion detection speed was achieved by choosing pyramid computation that was available in MATLAB \textit{imregtform} function.

%(Evidence of the feasibility)

Both Fig. \ref{fig:B_mode_demo} and \ref{fig:B_mode_testing}, along with the supplementary video, demonstrate how large motions can cause a target to move out of the imaging FoV. These results also effectively illustrate the use of robot-assisted motion tracking in maintaining the target within the imaging FoV, even under conditions of significant motion. The real-time performance of the motion tracking was evaluated by analysing the target moving distance in the imaging FoV during offline processing. With laptop-controlled motorised translation stages, the tracking performance was evaluated at five different speed levels and the evaluation was repeated twice. Target moving distance remained in the images increased with the moving speed of phantom, because the phantom was moving and the probe was still when doing the real-time data processing and faster speed moves more during the data processing.  As described in the introduction, image pattern changed with its position in the imaging FoV, especially for the unevenly distributed elements of the used matrix array. The change of image pattern affected the convergence of iterations in the image registration and might be the main reason of oscillations in the curves in Fig. \ref{fig:residual_motion_curve}. With current implementation in this study, the mean residual moving distances under the fastest testing trials were no more than 1.2 mm, as shown in Fig. \ref{fig:residual_motion_mean}. The residual motions for the target moving faster than the tested range could be preliminarily inferred by the regression in Fig. \ref{fig:residual_motion_mean}(d). Power Doppler images in Fig. \ref{fig:power_doppler_images} present coincident results with Fig. \ref{fig:B_mode_demo} that signals were more overlapping in the imaging FoV when using the robot-assisted real-time tracking. Image reconstructed in offline could reach super resolution with the motion tracking even if the imaging target moved in a distance twice the aperture size, demonstrating the feasibility of our proposed strategy to achieve SR imaging under large motions and the potentials its application in \textit{in vivo}.

%%(Future work)
While this work demonstrates significant progress in robot-assisted tracking for high-frame-rate ULM, several limitations should be noted as below.

    Validation on Animals: In this study, we only evaluated the technique on a lab phantom which consists of some artificial landmarks. The current study has not yet been validated on animals where respiration motion significantly affects probe positioning. Future work should focus on testing the system in such scenarios to assess its performance.% under more complicated conditions. % which might require better motion detection and robotic control methods.

    Data Processing and Memory Constraints: The frame rate and duration for continuous data acquisition were empirically chosen to avoid exceeding memory limits. \textcolor{black}{The single plane wave was used in this study for the consideration of least memory and computation cost, which results in lower-resolution and lower-SNR than compounding waves and might affect both real-time motion tracking and MB localization accuracy. Optimising the number of compounding waves to achieve a better balance between computational cost and image quality requires further investigation.}  Future improvements could include increasing memory capacity or implementing methods to offload data to disk in real-time, thereby overcoming current limitations related to data size.

    Signal Processing Limitations: The study utilised Amplitude Modulation (AM) for acquiring MB signals due to concerns that large tissue motions could impact the performance of Singular Value Decomposition (SVD) for clutter filtering\cite{demene2015spatiotemporal}. \textcolor{black}{However, tissue signal cancellation performance of AM or Pulse Inversion (PI) techniques can also be worse with faster target speed due to more signal phase shifts. Besides, SNR of contrast signals can be reduced with narrower bandwidth and with higher-frequency ultrasound due to decreased nonlinear components in MB signals\cite{tang2006nonlinear}. Low SNR and contrast can make  MB localization and tracking more challenging.} Future research should explore enhancing SVD performance for large motions \cite{solomon2019robustPCA,wahyulaksana2023higher} or developing motion correction techniques that preserve signal phases before applying SVD.

    Motion Detection Speed and Robustness: \textcolor{black}{When using the trick of reducing image volume for detection, a compromise between computation speed and maximum target distance between two frames has to be considered, which determines the maximum motion speed the method can accommodate.} The system currently operates at a period over 0.1 second, which may not be sufficient for all applications. \textcolor{black}{The motion detection problem is also simplified in this study by estimating motion of a single hyperechoic target via a MATLAB built-in algorithm}. Advancing online motion detection methods, either by developing new approaches or adapting existing offline methods \cite{de2018evaluation}, could improve real-time tracking performance, \textcolor{black}{guarantee the performance in complicated scenarios,} and reduce offline motion correction difficulties.
    
    \textcolor{black}{Elastic Deformation Manipulation: Since the probe is moved by the robot in a rigid space, this technique can only mitigate rigid motion through rigid motion tracking and cannot address the elastic deformation of the target within the FoV.  These investigation will focus on how to deal with elastic deformation for robot control: 1) analysing the impact of elastic deformation on rigid motion detection; and 2) developing a method that allows motion tracking in a rigid space while accounting for target deformation.}

    Robotic Control: The current implementation uses a basic robotic control law \textcolor{black}{and aims to keep moving target in FoV without axial probe motion considered}. To enhance real-time motion tracking, advanced control methods should be considered. For instance, the presence of air gaps between the probe and skin, or the impact of respiration on the probe’s movement, could be addressed with a hybrid force and position control strategy\cite{pierrot1999hippocrate,gilbertson2015force,ma2021novel}. Implementing force control in the depth direction and position control in lateral and elevational directions could improve \textcolor{black}{accuracy and safety of tracking}. Additionally, exploring robust control techniques such as active disturbance rejection control \cite{han2009pid} could further enhance the system's robustness.
    
    Surface and Path Considerations: The non-flat surface of the abdomen and respiratory motion could affect probe positioning. Addressing these challenges might involve generating tracking paths for imaging organs under ribs and implementing advanced control strategies to account for varying surface conditions.
    
\section{Conclusion}
This work developed and implemented a real-time robot-assisted motion tracking strategy designed to manage large target motions during ultrasound localization microscopy (ULM) data acquisition. It is the first feasibility study to demonstrate the application of a robot for tracking arbitrary target motions during high-frame-rate ULM data acquisition. The results highlighted the strategy's effectiveness in real-time tracking of target motion across multiple directions and confirmed the feasibility of achieving super-resolution imaging through offline processing. This approach not only addresses current challenges but also offers flexibility for future improvements. It shows great potential for handling respiration motion in clinical ULM imaging, paving the way for more accurate and reliable imaging in medical applications

\section{Acknowledge}
We would like to thank Prof. Yuanyi Zheng from Shanghai Sixth People's Hospital affiliated with Shanghai Jiao Tong University and Zhongliang Jiang from Technical University of Munich for their insightful discussions. We also would like to thank Cecilia G F Dunsterville and Krish K Desai for their help in acquiring data.
\bibliographystyle{IEEEtran}
\bibliography{sample}

\end{document}